\documentclass{webofc}
\usepackage[varg]{txfonts}   

\newcommand{\snn}{$\sqrt{s_{\mathrm{NN}}}~$}
\newcommand{\kst  }{$\mathrm{K^{*0}}~$}
\newcommand{\ks}{$\mathrm{K^{0}_{S}}~$}
\newcommand{\pt}{$p_{\mathrm{T}}$}

\begin{document}
\title{ Measurements of spin alignment of vector mesons and global polarization of hyperons with ALICE at the LHC}
\author{\firstname{Bedangadas} \lastname{Mohanty (for the ALICE collaboration)}\inst{1}\fnsep\thanks{\email{bedanga@niser.ac.in}} 
}
\institute{National Institute of Science Education and Research, HBNI, Jatni 752050, India
}

\abstract{We present the measurements related to global polarization of $\Lambda$ hyperons and 
spin alignment of  \kst vector mesons at mid-rapidity for Pb-Pb collisions at \snn= 2.76 TeV using the 
ALICE detector at the LHC. The global polarization measurements are carried out with respect to the first 
order event plane while the spin alignment measurements are carried out with respect to the 
production plane. No global polarization signal for $\Lambda$ is observed for 5-15\% and 15-50\% 
central Pb-Pb collisions. The spin density matrix element $\rho_{00}$ is found to have values slightly 
below 1/3 at low transverse momentum (\pt) for \kst   mesons, while it is consistent with 1/3 
(no spin alignment) at higher  \pt. No spin alignment is observed for \kst   in pp collisions 
at $\sqrt{s}$ = 13 TeV and for the spin zero hadron \ks in 20-40\% Pb-Pb collisions at \snn = 2.76 TeV.
}

\maketitle
\section{Introduction}
\label{sec:intro}
Large magnetic field~\cite{Kharzeev:2007jp} and large angular momentum~\cite{Becattini:2007sr} are 
expected to be created in the initial stages of high energy heavy-ion collisions. One of the physics 
interests of the heavy-ion program using the ALICE detector at the LHC is to look for signatures of these  
effects which can be looked into by studying the angular distributions 
of the decay daughters of hyperons and vector mesons~\cite{Abelev:2007zk,STAR:2017ckg,Abelev:2008ag}.  
The angular distributions are measured with respect to a quantization axis, which can either be perpendicular 
to the production plane of the hadron, or normal to the reaction plane of the system. The production plane is 
defined by the momentum of the hadron under study and the beam direction, whereas the reaction plane is 
defined by the impact parameter and beam direction. The angle denoted as $\theta^{*}$ is that 
made by one of the decay daughters of the hadron in the rest frame of the hadron with respect to the quantization 
axis. In general, the angular  distribution for vector mesons is expressed as ~\cite{Schilling:1969um},
\begin{equation}
\frac{\mathrm{d}N}{\mathrm{d}\cos{\theta^{*}}} = N_{0} [ 1 - \rho_{00} + \cos^{2}{\theta^{*}}(3\rho_{00} - 1 ) ],
\label{eqn1}
\end{equation}
while that for hyperons as~\cite{Abelev:2007zk},
\begin{equation}
\frac{\mathrm{d}N}{\mathrm{d}\cos{\theta^{*}}} =  \frac{1}{2} [ 1 + \alpha_{\mathrm{H}} P_{\mathrm{H}} \cos{\theta^{*}} ].
\label{eqn2}
\end{equation}
$N_{0}$ is a normalization constant, $\rho_{00}$ is the zeroth element of the spin density matrix, $\alpha_{\mathrm{H}}$ is the decay parameter and $P_{\mathrm{H}}$ is the polarization.
If initial conditions or the final hadronization process cause polarization effects in heavy-ion collisions, 
then the angular distributions as defined in Eq.\ref{eqn1} and Eq.\ref{eqn2} would become non uniform. 
This would lead to $\rho_{00}$ values being different from 1/3 and/or non-zero values for $P_{\mathrm{H}}$. 

In this work we present the first results at LHC energies related to the spin alignment of  \kst   vector 
mesons through the measurement of $\rho_{00}$ in pp and Pb-Pb collisions with respect to the production 
plane. In addition, we report on the global  polarization of $\Lambda$ hyperons through measurement of $P_{\mathrm{H}}$ in Pb-Pb collisions with respect to the event plane using the ALICE detector~\cite{Abelev:2014ffa}. 
\section{Global polarization of $\Lambda$ hyperons}
\label{sec:gp}
The global polarization studies of $\Lambda$ hyperons are carried out using 49 million minimum bias 
Pb-Pb events at \snn = 2.76 TeV. The analysis is performed at mid-rapidity (-0.5 $<$ $y$ $<$ 0.5) for 5-15\% 
and 15-50\% central collisions. The trigger and centrality selection use the V0 detectors~\cite{Abelev:2014ffa}. 
Flow based techniques are used as discussed in ~\cite{STAR:2017ckg,Timmins:2017gen}. The information from 
the Time Projection Chamber (TPC) is used to reconstruct the 
$\Lambda$ hyperons from their p and $\pi$ decay channels. The measurements are done using the first 
order event plane reconstructed using the zero degree calorimeter~\cite{Timmins:2017gen}.  The \pt~
integrated values of $P_{\mathrm{H}}$ for $\Lambda$ hyperons in -0.5 $<$ $y$ $<$ 0.5 are measured 
to be $P_{\mathrm{H}}(\Lambda)$ = -0.01 $\pm$ 0.13 (stat) $\pm$ 0.04 (syst) and $P_{\mathrm{H}}(\Lambda)$ 
= -0.08 $\pm$ 0.10 (stat) $\pm$ 0.04 (syst) for 5-15\% and 15-50\% central Pb--Pb collisions at \snn = 2.76 TeV, 
respectively~\cite{Timmins:2017gen}. The corresponding values for  $\bar{\Lambda}$ are 
$P_{\mathrm{H}}(\bar{\Lambda})$ = -0.09 $\pm$ 0.13 (stat) $\pm$ 0.08 (syst) for 5-15\% and 
$P_{\mathrm{H}}(\bar{\Lambda})$ = 0.05 $\pm$ 0.10 (stat) $\pm$ 0.03 (syst) for 15-50\% central Pb--Pb 
collisions~\cite{Timmins:2017gen}. The $P_{\mathrm{H}}$ values are consistent with zero.
\section{Spin alignment of \kst vector mesons}
\label{sec:sp}
\begin{figure}[h]
\centering 
\includegraphics[scale=0.25]{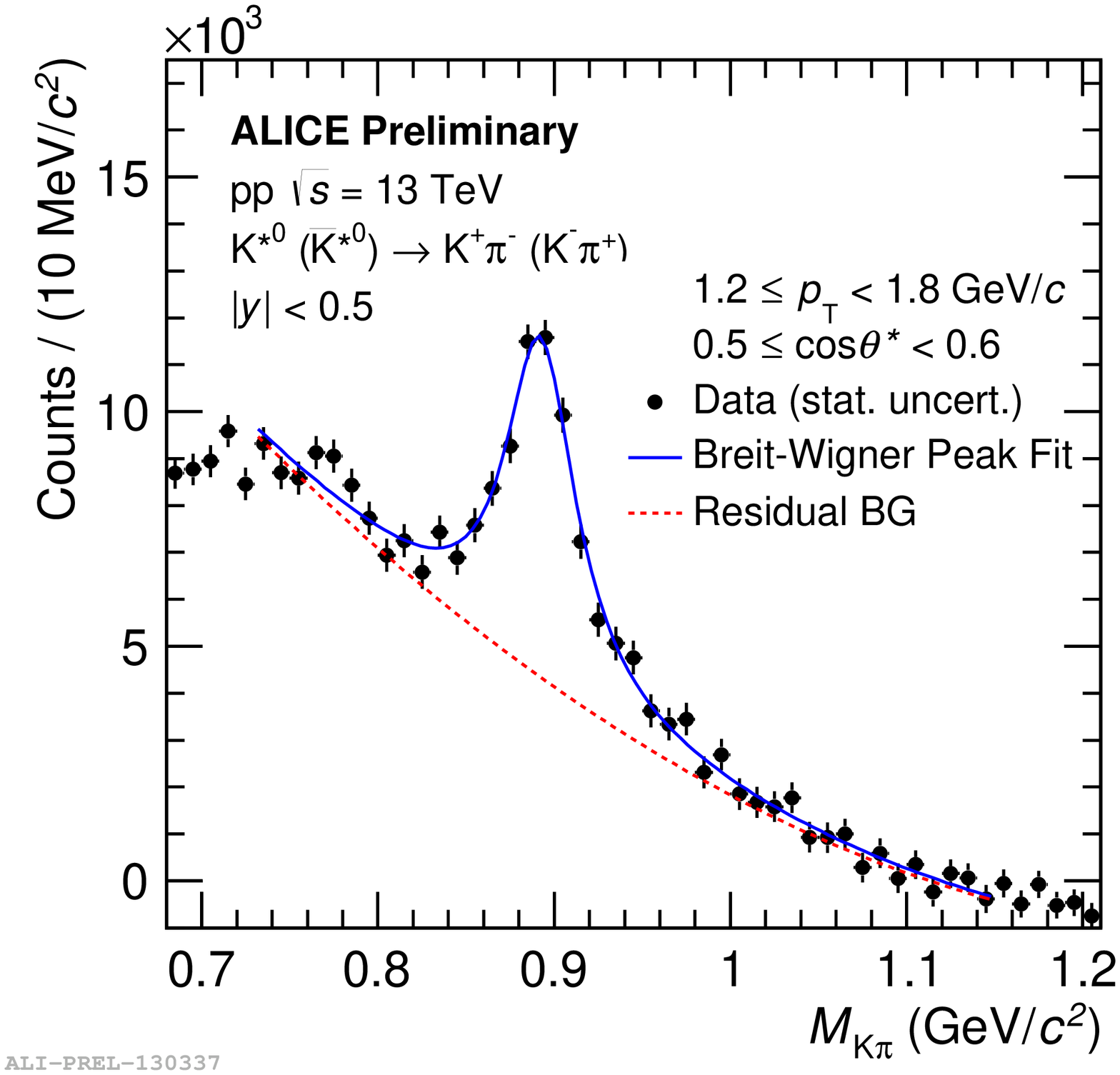}
\includegraphics[scale=0.25]{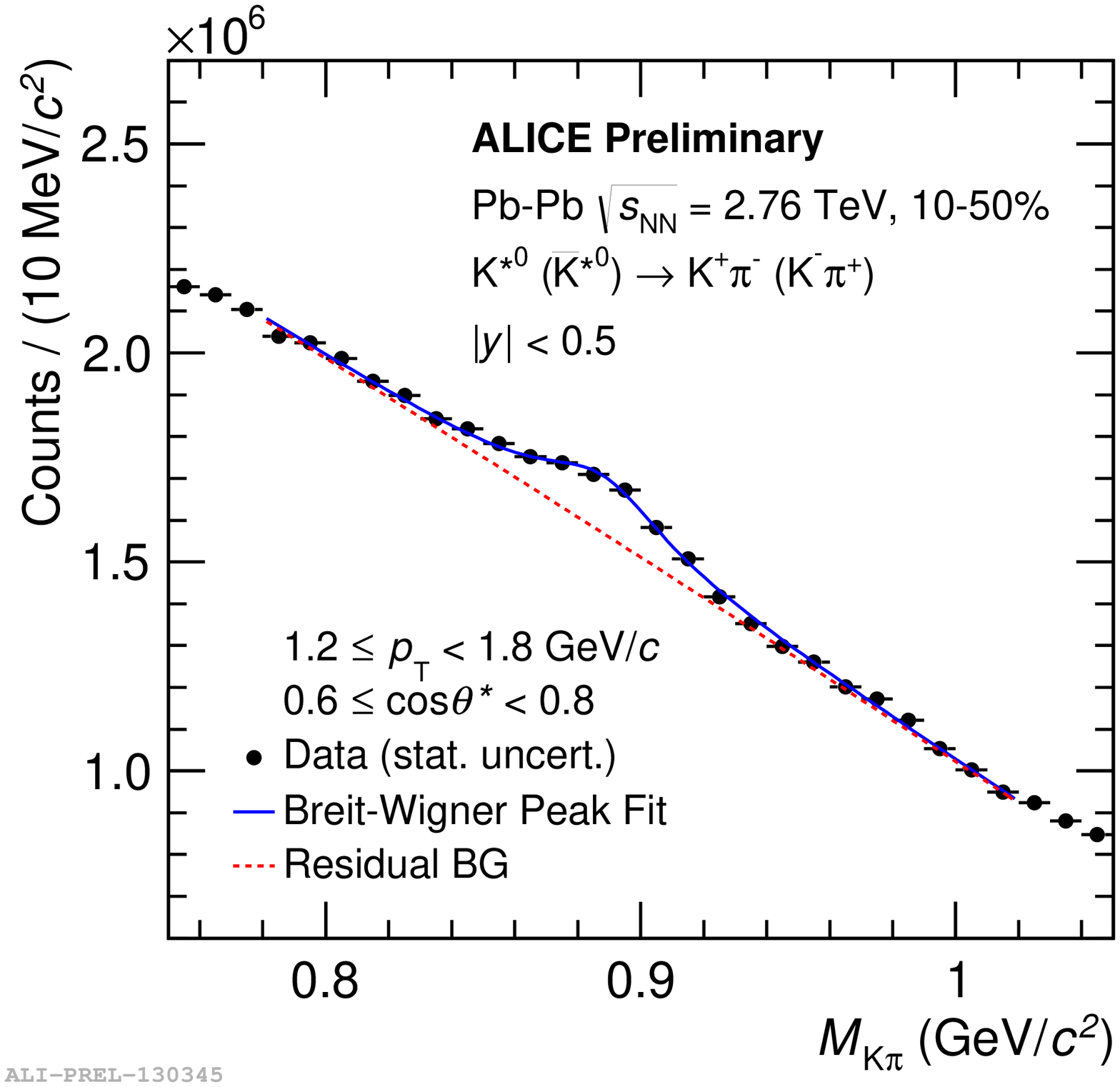}
\caption{(Color online) Left Panel: Kaon-Pion invariant mass ($M_{\mathrm{K\pi}}$) distribution at 
mid-rapidity after mixed background subtraction in minimum bias pp collisions at $\sqrt{s}$ = 13 TeV. 
The data are fitted with a Breit-Wigner function and a $2^{nd}$ order polynomial function in $M_{\mathrm{K\pi}}$ 
reflecting the residual background. Right Panel: Same as left panel but for 10-50\% central Pb--Pb collisions 
at \snn = 2.76 TeV. }
\label{inv} 
\end{figure}
\begin{figure}[h]
\centering 
\includegraphics[scale=0.28]{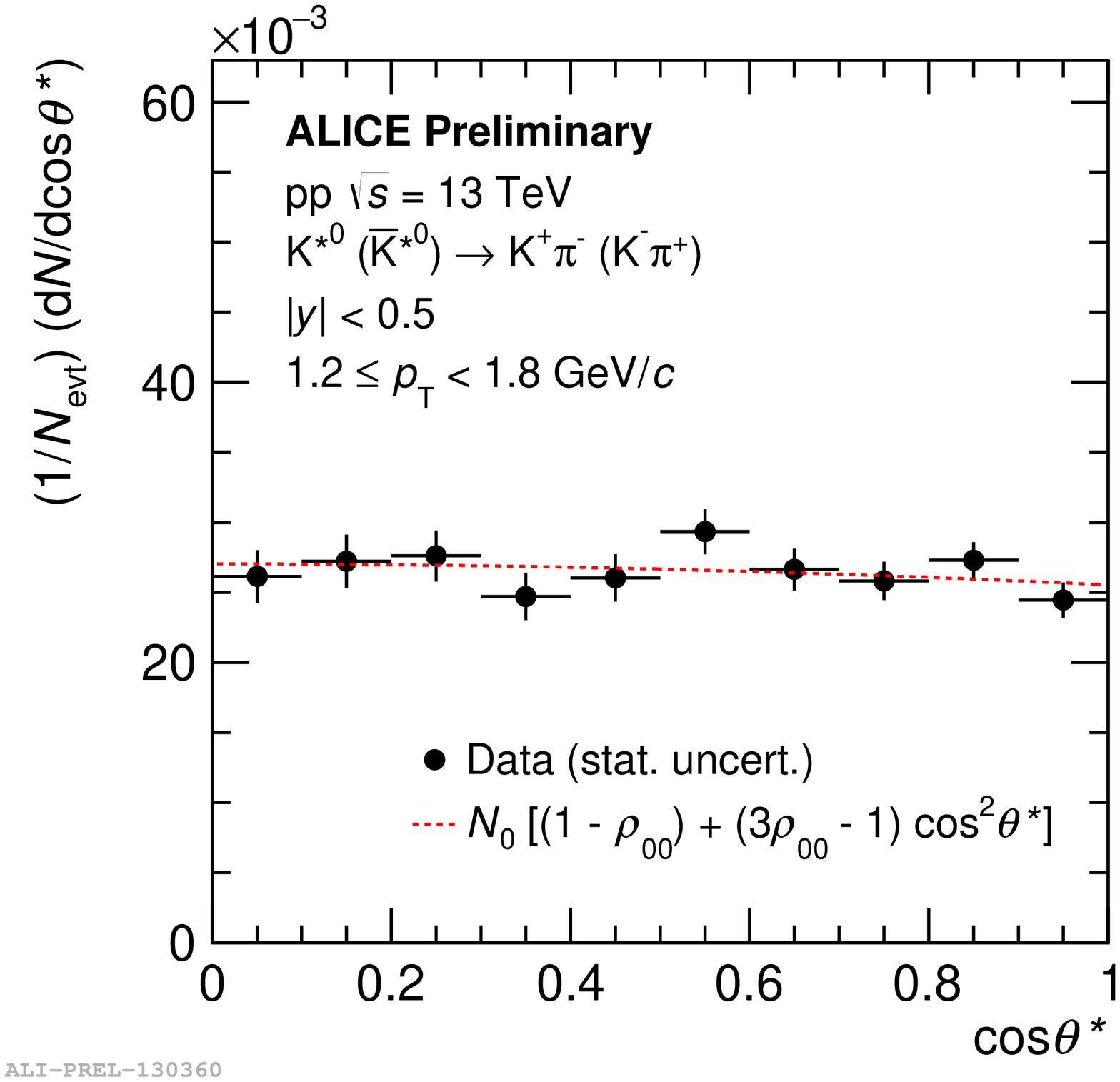}
\includegraphics[scale=0.28]{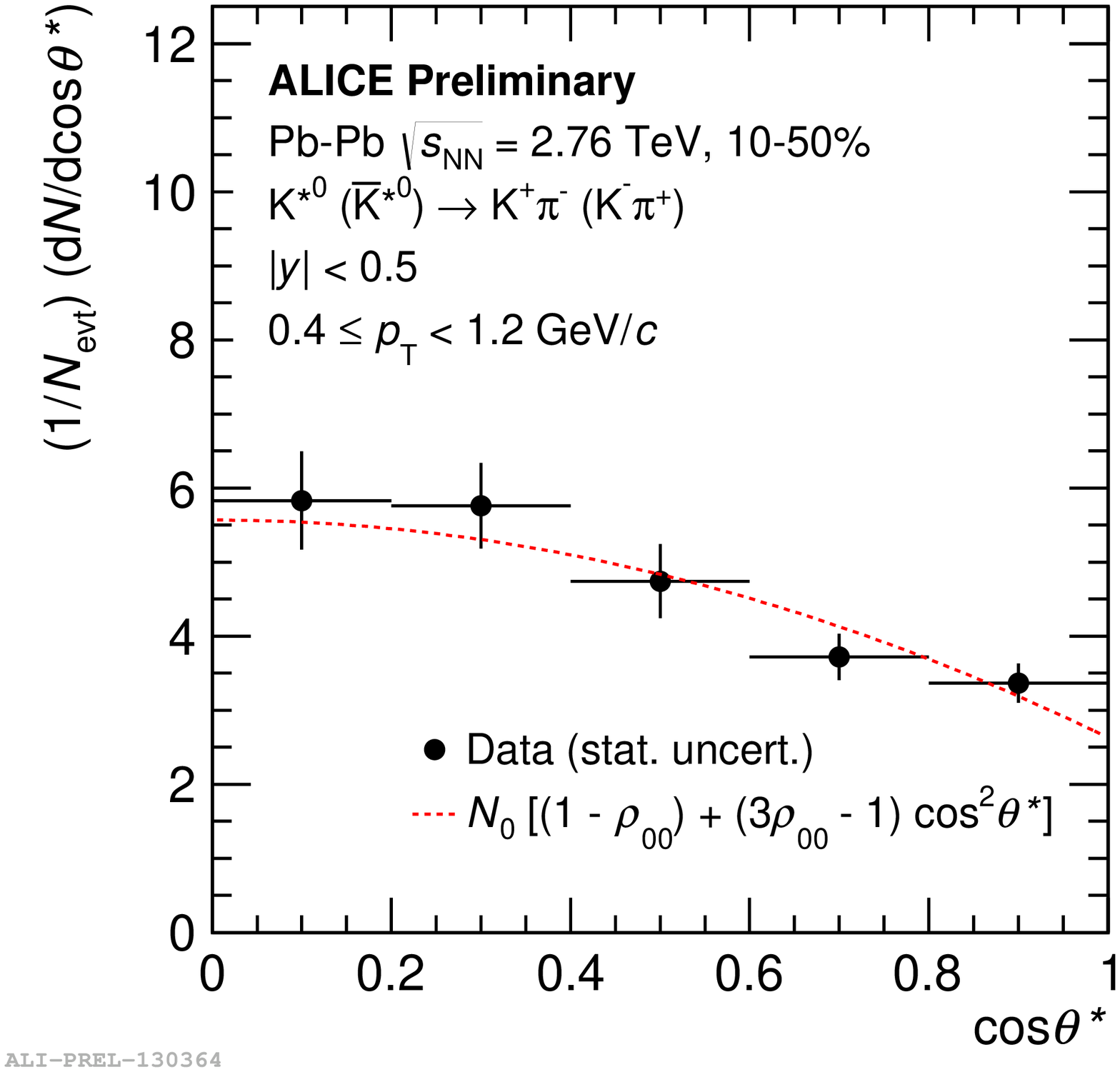}
\caption{(Color online) Left Panel: $\mathrm{d}N/\mathrm{d}\cos{\theta^{*}}$ versus $\cos{\theta^{*}}$ 
at mid-rapidity for minimum bias pp collisions at $\sqrt{s}$ = 13 TeV. Right Panel: Same but for 
10-50\% central Pb--Pb collisions at \snn = 2.76 TeV. The error bars shown are statistical only. 
The angular distributions are obtained with respect to production plane.}
\label{dist} 
\end{figure}

The spin alignment studies for the spin one meson \kst  are carried out using 43 million minimum bias events in pp 
collisions at $\sqrt{s}$ = 13 TeV and 14 million minimum bias events in Pb--Pb collisions at \snn = 2.76 TeV. 
The \kst analysis in Pb--Pb collisions uses the data taken in year the 2010 whereas the global polarization 
studies of $\Lambda$ hyperons use both the data set taken in the year 2010 and 2011.
The measurements are performed at mid-rapidity for both pp and Pb--Pb collisions. Results for 10-50\% collision 
centrality in the Pb--Pb system are presented here. In order to study the null hypothesis, a similar analysis 
for 20-40\% collision centrality using the spin zero meson \ks is also carried out. The \kst  
are reconstructed via the decay into charged K and $\pi$ while the \ks is reconstructed via its decay into $\pi\pi$. The ALICE TPC and TOF are used for identifying the decay products of the above hadrons and the signals are extracted via the invariant mass technique~\cite{Adam:2017zbf,Abelev:2014uua}. The 
background, dominantly coming from combinatorics, is removed by constructing the invariant mass 
distributions using mixed event techniques~\cite{Adam:2017zbf,Abelev:2014uua}.  The mixed event 
background subtracted invariant mass $M_{\mathrm{K\pi}}$ distributions for pp and Pb--Pb collisions are 
shown in Fig.~\ref{inv}. The results are shown for a particular \pt ~and $\cos{\theta^{*}}$ bin 
as a representative case, but the same procedure is followed for each \pt~and $\cos{\theta^{*}}$ bin of the analysis.
 The distributions are then fitted with a Breit-Wigner function for the signal and $2^{nd}$ order polynomial for the residual background to extract the yields, which
are then corrected for the corresponding  
reconstruction efficiency and acceptance  in each $\cos{\theta^{*}}$ and \pt~ bin using Monte Carlo simulations of the ALICE detector and response ~\cite{Adam:2017zbf,Abelev:2014uua}. The efficiency and acceptance 
corrected $\mathrm{d}N/\mathrm{d}\cos{\theta^{*}}$ distributions at mid-rapidity for 1.2 $\le$ \pt~$<$ 1.8 GeV/$c$ 
in minimum bias pp collisions and for 0.4 $\le$ \pt~$<$ 1.2 GeV/$c$ in 10-50\% central Pb-Pb 
collisions are shown in Fig.~\ref{dist}. These distributions are fitted with the functional form 
shown in Eq.~1 to extract the $\rho_{00}$ values for each \pt ~bin. 
\begin{figure}[h]
\centering
\includegraphics[scale=0.26]{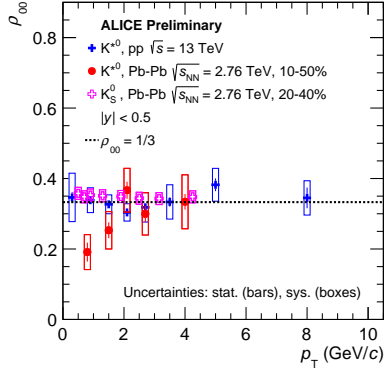}
\caption{(Color online) $\rho_{00}$ versus \pt~extracted from the $\cos{\theta^{*}}$ distributions at mid-rapidity for \kst  vector meson in minimum bias pp collisions at  $\sqrt{s}$ = 13 TeV and 10-50\% central Pb--Pb  collisions at \snn = 2.76 TeV. Also shown are the corresponding results for \ks spin zero meson for 20-40\% Pb--Pb collisions at \snn = 2.76 TeV. The vertical lines on the data points are statistical errors and the boxes reflect the systematic uncertainties on $\rho_{00}$.}
\label{rho}
\end{figure}
Figure~\ref{rho} shows the $\rho_{00}$ values as a function of \pt~for \kst  and $\mathrm{K^{0}_{S}}$. The \kst  results for pp collisions at  $\sqrt{s}$ = 13 TeV  are consistent with no spin alignment ($\rho_{00}$ $\sim$ 1/3). The $\rho_{00}$ values for \kst   at 0.4 $\leq$ \pt ~$<$ 1.2 GeV/$c$ is about 2.5 standard deviations ($\sigma$) lower than 1/3, about 1.4$\sigma$ lower than 1/3 for the \pt~bin 1.2 - 1.8 GeV/$\it{c}$ and consistent with 1/3 for higher \pt~values in 10-50\% central Pb--Pb  collisions at \snn = 2.76 TeV. The $\sigma$ is calculated by adding the statistical and systematic uncertainties on $\rho_{00}$  in quadrature. The $\rho_{00}$ values are consistent with 1/3 for the whole \pt~range studied for \ks in 20-40\% central Pb--Pb  collisions at \snn = 2.76 TeV. The sources of systematic uncertainties on $\rho_{00}$  include: reconstruction efficiency and acceptance, track selection conditions, particle identification procedure, signal extraction procedure, and material budget estimates ~\cite{Adam:2017zbf,Abelev:2014uua}. 
\section{Summary and outlook}
\label{sec:summary}
We have presented results related to the global polarization of $\Lambda$ hyperons measured with respect to the reaction plane and those related to spin alignment of \kst with respect to the production plane in Pb-Pb collisions at \snn = 2.76 TeV. No polarization effect was observed for the $\Lambda$ produced at mid-rapidity for 5-15\% and 15-50\% central collisions. The spin density matrix element $\rho_{00}$ extracted from the angular distributions of the decay daughter of \kst at mid-rapidity in 10-50\% central Pb-Pb collisions was found to be slightly lower than 1/3 for \pt~below 1.8 GeV/$c$. The  $\rho_{00}$ values are consistent with 1/3, namely indicating no spin alignment, for high \pt~\kst  in Pb-Pb collisions, for the full \pt~range studied in pp collisions and for spin zero \ks produced in 20-40\% central Pb-Pb collisions. 

The analysis of the centrality dependence of spin alignment for vector mesons with respect to both production and reaction planes is ongoing. Results with increased statistical precision are expected for both global polarization and spin alignment studies with the Pb-Pb data set at \snn = 5.02 TeV. This will help in clarifying the significance of the observed effects.


\end{document}